\documentstyle[referee,epsfig]{mn}

\def\fl{S_{\rm 1.4 \, GHz}}

\begin{document}

\title{Quantifying angular clustering in wide-area radio surveys}

\author[Chris Blake and Jasper Wall]{Chris Blake\footnotemark \, and
Jasper Wall \\ Astrophysics, University of Oxford, Keble Road, Oxford,
OX1 3RH}

\maketitle

\begin{abstract}
We quantify the angular clustering of radio galaxies in the NVSS and
FIRST surveys using the two-point correlation function and the moments
of counts-in-cells -- both important points of comparison with theory.
These investigations consistently demonstrate that the slope of the
correlation function for radio galaxies agrees with that for
optically-selected galaxies, $\gamma \approx 1.8$.  We describe how to
disentangle the imprint of galaxy clustering from the two
observational problems: resolution of radio galaxies into multiple
components and gradients in source surface density induced by
difficulties in processing ``snapshot'' radio observations
(significant in both surveys below $\fl \sim 15$ mJy).  This study
disagrees in some respects with previous analyses of the angular
clustering of radio galaxies.
\end{abstract}

\section{Introduction}
\renewcommand{\thefootnote}{\fnsymbol{footnote}}
\setcounter{footnote}{1}
\footnotetext{E-mail: cab@astro.ox.ac.uk}

Describing the large-scale structure of the Universe is of fundamental
importance for testing theories of galaxy and structure formation and
for measuring the cosmological parameters.  The largest structures
require delineation by the deepest, widest surveys, currently
represented by surveys for radio AGN.  These contain objects to
redshifts of at least $z \sim 4$: the radio emission marking these
objects is not affected by dust obscuration, large-scale calibration
effects should be minimal, and the number of objects in the current
generation of radio surveys such as WENSS, FIRST and NVSS reach $\sim
10^6$ over substantial fractions of the sky.

We see the distribution of galaxies projected on the sky, but this is
still useful to quantify: it is easy to assemble a large sample of
objects and the angular clustering can be de-projected (in a global
statistical manner) to measure the spatial clustering, conclusions
being reached in the absence of complete redshift information.  There
are many sophisticated methods for quantifying the angular
distribution of galaxies.  These include spherical harmonic analysis
(e.g. Baleisis et al. 1998), percolation analysis (Bhavsar \& Barrow
1983) and minimal spanning trees (Krzewina \& Saslaw 1996).  Chiang \&
Coles (2000) emphasize the importance of maintaining the phase
information of the clustering for describing morphology.  In contrast,
here we use two of the crudest statistics for describing angular
structure: the two-point angular correlation function and the moments
of counts-in-cells.  It is well-known that these methods lose much of
the clustering information: two very different distributions can have
the same two-point correlation function.  However, these statistics
are simple to interpret and hence reveal the fundamental observational
problems and survey limitations.  They provide simple points of
contact with prediction, have well-understood statistical errors and
together provide a consistency check.  They must be understood and
must give consistent results before application of more powerful
techniques can be considered.

Correlation function analyses (Peebles 1980), widely used since the
early days of clustering investigations, have been extensively applied
in the optical regime, for example to the APM survey (Maddox et
al. 1996).  Here, the correlation function typically shows a power-law
behaviour $w(\theta) \propto \theta^{1-\gamma}$ with $\gamma \approx
1.8$ on small scales, with a steepening break to larger scales.  The
key difference between angular correlation function analyses in the
optical and radio regimes is in the latter, the wide redshift range of
radio sources washes out much of the clustering amplitude through the
superposition of unrelated redshift slices.  Hence an angular
clustering signal has only been measurable in the most recent radio
surveys, initially with marginal detections in the Green Bank 87GB
survey (Kooiman et al. 1995) and the Parkes-MIT-NRAO (PMN) survey
(Loan et al. 1997).

The latest generation of deep radio surveys -- FIRST (Becker et
al. 1995), WENSS (Rengelink et al. 1998) and NVSS (Condon et al. 1998)
-- reveal the imprint of structure more clearly.  The correlation
function has been measured for WENSS by Rengelink et al. (1998), and
for FIRST by Cress et al. (1996) and, in a pioneering and innovative
series of papers, by Magliocchetti et al. (1998).  These studies
concluded that the slope of the correlation function for radio
galaxies was steep ($\gamma > 2$).  Cress \& Kamionkowski (1998) and
Magliocchetti et al. (1999) modelled the 3D clustering from these
analyses, including the behaviour of bias with epoch.  Magliocchetti
et al. (1998) also carried out a counts-in-cells analysis of the FIRST
survey, detecting significant skewness.

These results motivated our present investigation.  The NVSS had never
been investigated for large-scale structure effects, and we wished to
determine if the more extensive sky coverage and source list of $\sim
2 \times 10^6$ objects led to conclusions compatible with FIRST, with
higher signal-to-noise ratio affording further insight.  We wished to
understand how robust the conclusions from FIRST were, given the issue
of over-resolution and the consequent need to ``combine'' catalogue
sources from multiple-component radio galaxies.  We wished to examine
the compatibility of results from counts-in-cells and
correlation-function analyses.  With these aims in mind, NVSS and
FIRST, at the same frequency but at resolutions differing by a factor
of 9, suggest an ideal comparative study.  Our initial results
(measurement of the NVSS angular correlation function) were presented
in Blake \& Wall (2002).

To proceed we first describe the two surveys, NVSS and FIRST.  Section
\ref{secquan} summarizes the clustering statistics we use and Section
\ref{secobseff} discusses the observational issues bound to impact
upon large-scale structure analyses.  Sections \ref{seccorr} and
\ref{seccell} derive angular correlation functions and counts-in-cells
for each of NVSS and FIRST, and Section \ref{secconc} compiles the
conclusions.

\section{The radio surveys: NVSS and FIRST}

\subsection{NVSS}

\begin{figure}
\center
\epsfig{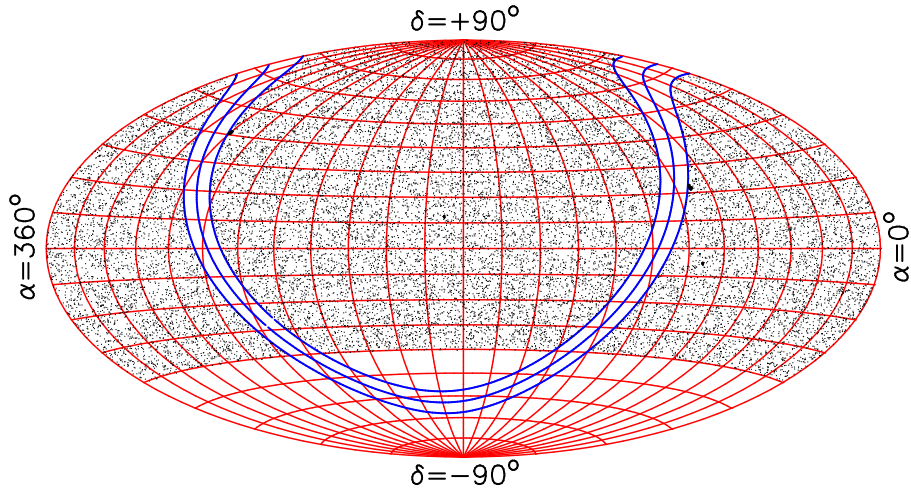}
\caption{NVSS catalogue entries with $\fl > 200$ mJy in an equal-area
projection.  The Galactic plane and Galactic latitudes $\pm 5^\circ$
are also plotted; sources within this region are masked from our
large-scale structure analysis as many are Galactic in origin.}
\label{fignvsssurv}
\end{figure}

The NVSS (NRAO VLA Sky Survey, Condon et al. 1998) was carried out
with the VLA at an observing frequency of 1.4 GHz over the period 1993
-- 1996 and covers the whole sky north of declination $-40^\circ$
(33,884 square degrees or 82 per cent of the celestial sphere).  The
source catalogue contains $1.8 \times 10^6$ sources and is claimed to
be 99 per cent complete at integrated flux density $\fl = 3.5$ mJy and
50 per cent complete at 2.5 mJy.  These figures are differential
completenesses, i.e. 99 per cent of all sources with $\fl = 3.5$ mJy
appear in the NVSS catalogue.  The survey was performed with the VLA
in D configuration, with DnC configuration used for fields at high
zenith angles ($\delta < -10^\circ, \delta > 78^\circ$), and the FWHM
of the synthesized beam is about 45 arc-seconds.  The raw fitted
source parameters are processed by a computer program provided by the
survey team called NVSSlist, which performs the deconvolution and
corrects for known biases to produce source diameters and integrated
flux densities.  Details are given by Condon et al. (1998); NVSSlist
version 2.16 (March 2001) was used for this investigation.  Figure
\ref{fignvsssurv} is a plot of NVSS catalogue entries with $\fl > 200$
mJy.

\subsection{FIRST}

The FIRST (Faint Images of the Radio Sky at Twenty centimetres)
survey, also carried out at an observing frequency of 1.4 GHz but with
the VLA in B configuration, began in 1993 and continues.  It covers
the North and South Galactic caps and the B configuration of the VLA
yields an angular resolution of about 5 arcsec.  Details of the survey
design, analysis and catalogue generation are given in Becker, White
\& Helfand (1995) and White et al. (1997).  These papers claim that
the survey is 95 per cent complete at $\fl = 2$ mJy and 80 per cent
complete at 1 mJy.  These figures are cumulative completenesses,
i.e. 95 per cent of sources with $\fl > 2$ mJy appear in the FIRST
catalogue.

The latest publicly-available catalogue dates from 15 October 2001.
It contains 771,076 sources and covers a total of 8565 square degrees
(7954 in the north galactic gap and 611 in the south), or 21 per cent
of the celestial sphere.  The raw catalogue contains a number of
spurious entries representing sidelobe responses from nearby brighter
sources.  As described in White et al. (1997), the FIRST survey team
developed an oblique decision-tree program to identify and flag these
sidelobes.  In the 15 October 2001 catalogue, 28,017 sources are
flagged and were excluded from our analysis.

\begin{figure}
\center
\epsfig{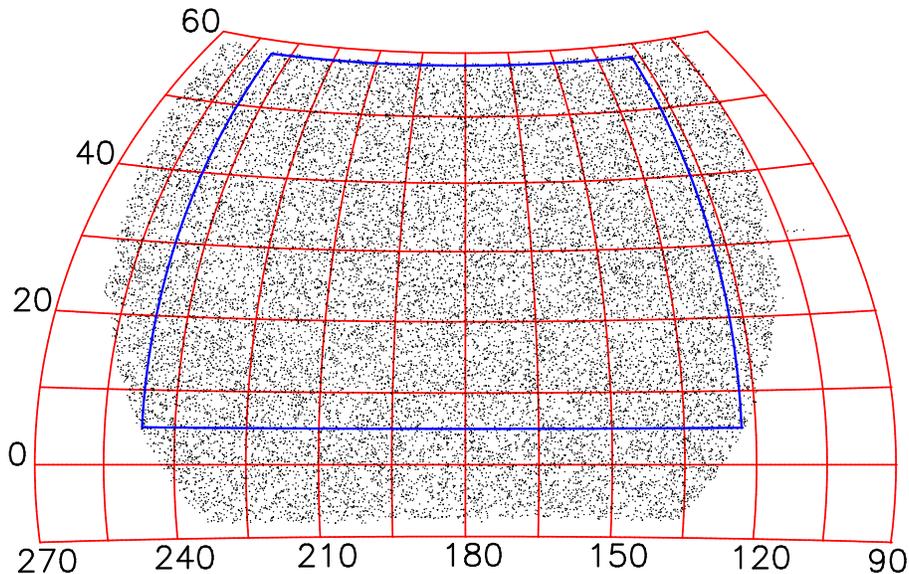}
\caption{FIRST catalogue entries with $\fl > 50$ mJy from the 15
October 2001 catalogue, plotted in an equal-area projection.  The
region for which our large-scale structure analysis was performed
($123^\circ < \alpha < 247^\circ$, $5^\circ < \delta < 58^\circ$) is
indicated.}
\label{figfirstsurv}
\end{figure}

In Figure \ref{figfirstsurv} we plot FIRST catalogue entries with $\fl
> 50$ mJy in the northern sky, and we outline the contiguous region
for which our large-scale structure analysis was performed ($123^\circ
< \alpha < 247^\circ$, $5^\circ < \delta < 58^\circ$).

\section{Quantifying the angular clustering of galaxies}
\label{secquan}

A common way of quantifying the clustering in an angular distribution
of galaxies is by using the {\it angular correlation function},
$w(\theta)$.  This compares the observed (clustered) distribution to a
random (unclustered) distribution of points across the same survey
area, by simply measuring the fractional increase in the number of
close pairs separated by angle $\theta$.  Specifically, if we let
$DD(\theta)$ be the number of unique pairs of galaxies with
separations $\theta \rightarrow \theta + \delta \theta$, and
$RR(\theta)$ be the number of random pairs in the same separation
range, then $w(\theta)$ can be estimated as $w(\theta) = (DD-RR)/RR$.
Further investigation reveals that the statistical error on $w$ may be
minimized by averaging over a large number of random sets ($RR
\rightarrow \overline{RR}$) and by using a different estimator for $w$
(see Landy \& Szalay 1993).  We adopt the Landy-Szalay estimator for
our investigations, as this has minimal statistical bias and variance.

Another simple way to quantify the galaxy distribution is to grid the
sky into cells of fixed area and shape, and count the number of
sources that lie in each cell.  {\it Counts-in-cells} yields the
probability distribution $P(N)$ of finding $N$ sources in a cell, and
the moments of the distribution such as the variance $\mu_2 =
\overline{(N - \overline{N})^2}$ and skewness $\mu_3 = \overline{(N -
\overline{N})^3}$ (the horizontal bar indicates an average over
cells).  A clustered distribution produces a higher variance than a
random distribution because cells may lie in clusters or voids,
broadening the probability distribution $P(N)$.  Skewness is important
because, assuming Gaussian primordial perturbations and linear theory,
the skewness of counts-in-cells remains zero (Peebles 1980).
Measurement of a non-zero skewness therefore indicates either
non-linear gravitational clustering or non-Gaussian initial
conditions.

Whereas the angular correlation function bins pair separations into
small intervals, a counts-in-cells analysis combines information from
a range of angular scales up to the cell size, effectively measuring
an average $w(\theta)$ (see equation \ref{eqy1}).  By avoiding the
binning of angular separations, the counts-in-cells is less affected
by ``shot noise'' and is a more sensitive probe of long-range
correlations.  A simple relation exists between $w(\theta)$ and
$\mu_2$ ($\S 6.1$), hence we can verify that they are consistent for a
given distribution.  Neither quantity provides a {\it complete
statistical description} of the galaxy distribution: this can be
encompassed by the hierarchy of correlation functions or the full
probability distribution $P(N)$; both are harder to interpret and to
compare with theory.  Nor are $w(\theta)$ and $\mu_2$ well-suited for
describing the {\it morphology} of the distribution (do galaxies
cluster in filaments, sheets or clumps?) or its {\it topology} (how do
the filaments or sheets join up to form the global pattern?).
However, these statistics are easy to measure, provide a simple point
of contact with prediction, and can reveal observational problems with
the survey data.

\section{Observational effects impacting on large-scale structure
studies}
\label{secobseff}

\subsection{Resolution effects}
\label{secres}

\begin{figure}
\center
\epsfig{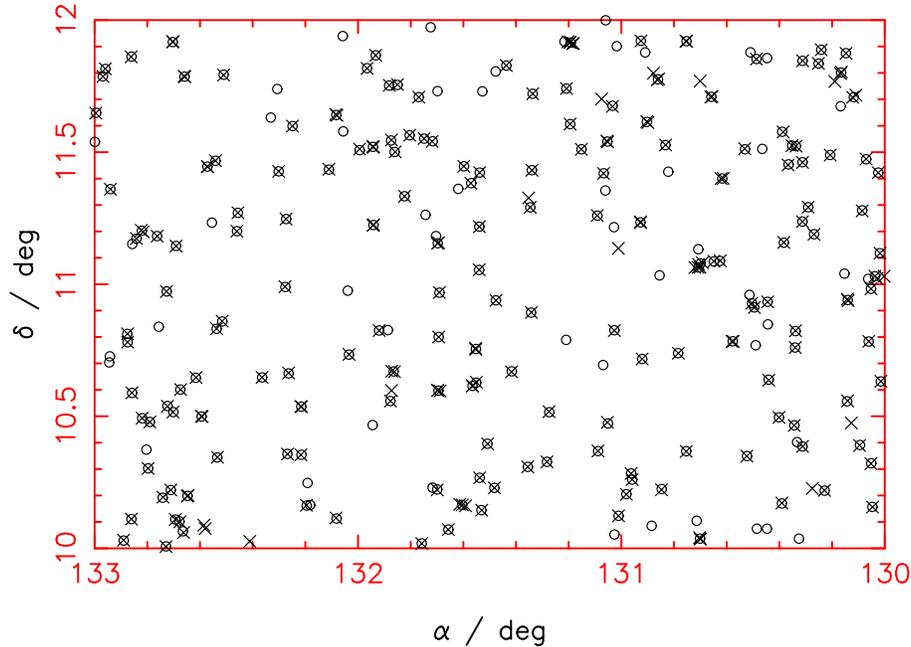}
\caption{NVSS catalogue entries (circles) and FIRST entries (crosses)
in a randomly-chosen region of sky common to each survey.  We only
plot sources with $\fl > 3.5$ mJy, above which threshold both surveys
are claimed to be complete.}
\label{figoverplot}
\end{figure}

The NVSS and FIRST surveys are at the same observing frequency (1.4
GHz) but at resolutions differing by a factor of 9, and therefore
provide an excellent comparative study of survey resolution effects on
the observed properties of radio galaxies.  In Figure
\ref{figoverplot} we plot NVSS and FIRST catalogue entries with
integrated flux densities $\fl > 3.5$ mJy (at which threshold both
surveys are claimed to be complete) in a randomly chosen $3^\circ
\times 2^\circ$ patch of sky.  This region contains 230 FIRST entries
and 228 NVSS entries.  We see that:
\begin{itemize}
\item Most radio sources are detected in both surveys.
\item Of the objects appearing in just one survey, many more appear in
NVSS than FIRST.
\item The surface densities are almost identical because many FIRST
objects are multiple components of the same galaxy.
\end{itemize}

These facts can be explained by the superior angular resolution of
FIRST, which picks up multiple radio components unresolved by NVSS.
But this high resolution provides FIRST with much poorer sensitivity
to surface brightness, losing flux from extended sources even well
above the survey limit.  Thus the objects appearing solely in NVSS in
Figure \ref{figoverplot} have fallen below the 3.5 mJy threshold in
FIRST.  The small number of sources which only appear in FIRST are due
to either statistical flux-density errors, source variability,
unrecognised sidelobes or noise spikes.

\begin{figure}
\center
\epsfig{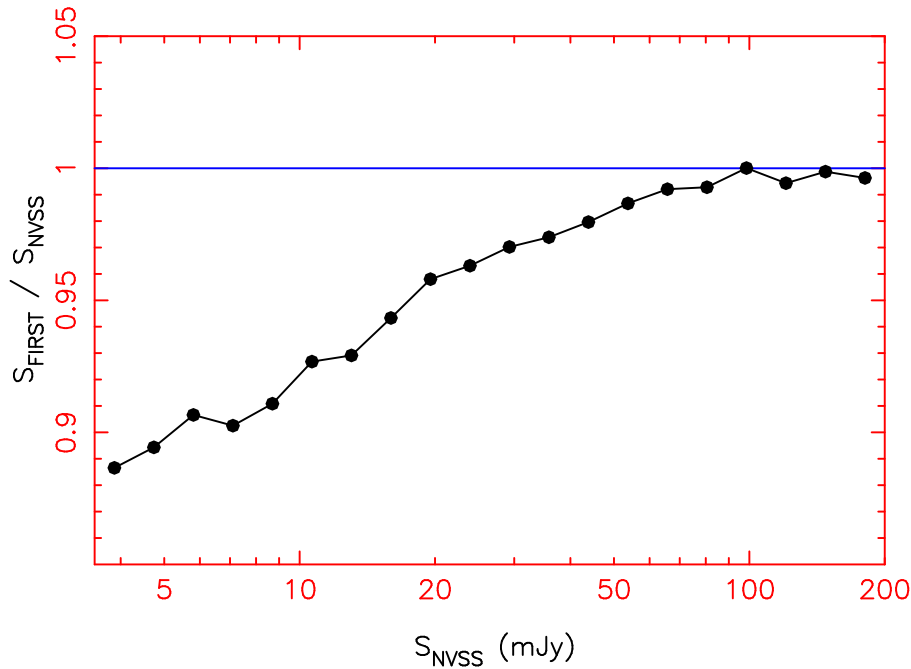}
\caption{The average ratio of FIRST to NVSS flux density in NVSS
flux-density bands for sources matched between the surveys (with
matching tolerance 10 arc-seconds).  The matched sources are chosen to
be isolated (having nearest neighbours more distant than 2 arcmin) in
both surveys to prevent confusion from galaxies resolved into multiple
components.}
\label{figfluxrat}
\end{figure}

The fact that FIRST under-estimates flux densities is confirmed by
Figure \ref{figfluxrat}, which plots the average ratio of FIRST to
NVSS integrated flux density for sources matched between the surveys
for different NVSS flux-density bands.  The matched sources are chosen
to be {\it isolated} in both surveys (having nearest neighbours more
distant than 2 arcmin) to prevent confusion from galaxies resolved as
multiple radio components.  Figure \ref{figfluxrat} shows that on
average FIRST loses 10 per cent of the flux density from 3.5 mJy
sources, a fraction which decreases with increasing flux density.

\begin{figure}
\center
\epsfig{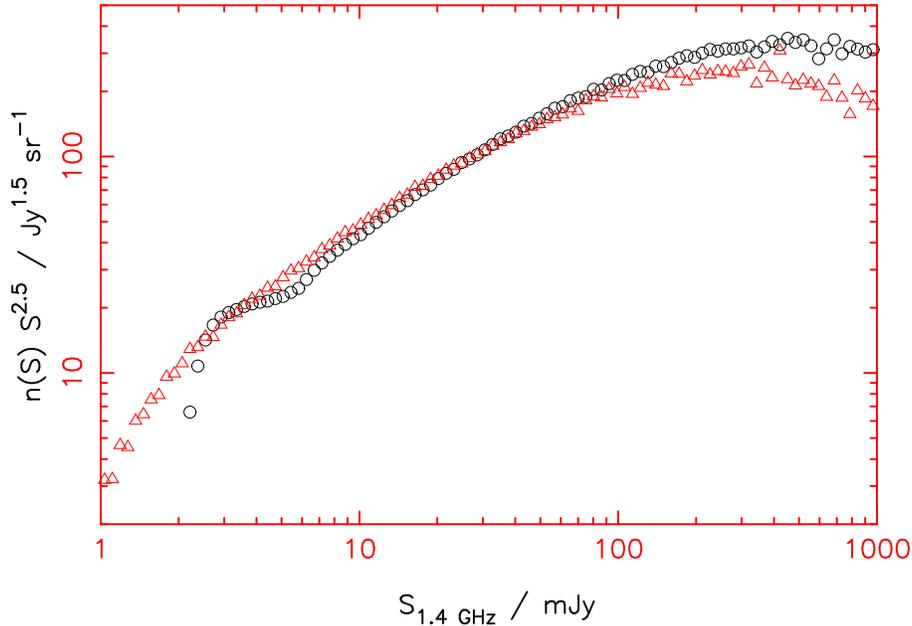}
\caption{Differential source counts for NVSS (circles) and FIRST
(triangles).  The counts are normalized to a Euclidean universe, so
that the $y$-axis plots $n(S) \times \, S^{2.5}$ where $n(S) \, dS$ is
the number of sources found per steradian in the flux-density range $S
\rightarrow S + dS$.  The error in each point is smaller than the
plotted symbol (except at very high fluxes).  At faint flux densities
both curves under-estimate the true source count due to incompleteness
effects.}
\label{figsrcount}
\end{figure}

In Figure \ref{figsrcount} we overplot differential source counts for
FIRST and NVSS for 1 mJy $< \fl <$ 1 Jy.  The curves are in rough
agreement.  However, the NVSS source count has an unphysical shape
below $\fl \approx 10$ mJy.  This distortion arises because NVSSlist
uses a different algorithm to convert raw fitted peak amplitudes to
integrated flux densities, depending on whether a source is classified
as extended or not (Condon et al. 1998).  In addition, a larger number
of bright ($> 100$ mJy) sources are detected in NVSS than in FIRST.
This is due to the finer angular resolution of FIRST breaking up
bright sources into more radio components.

These comparisons demonstrate the significant impact of survey
resolution and surface brightness sensitivity on the observed
properties of radio galaxies.  However, the effect of these flux
biases on the deduced large-scale clustering should be minimal: the
broadness of the radio galaxy luminosity function ensures that the
observed clustering is not a strong function of flux density above 3.5
mJy.

\subsection{Gradients in source surface density}
\label{secsurf}

\begin{figure}
\center
\epsfig{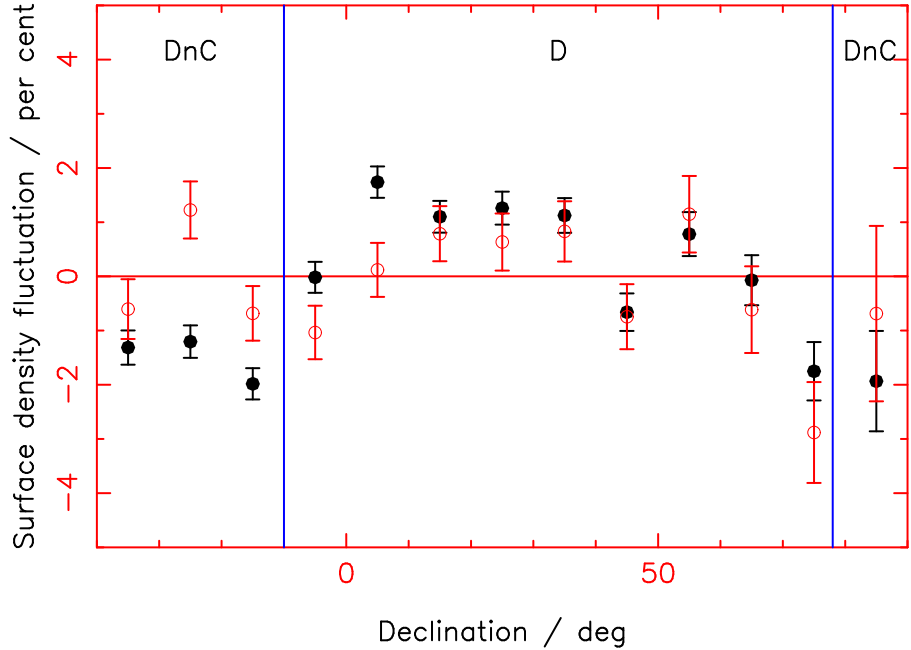}
\caption{Variations in NVSS source surface density as a function of
declination for sources with integrated flux densities above 3.5 mJy
(solid circles) and 15 mJy (open circles).  Sources have been binned
in declination bands of width $10^\circ$.  The declination range of
each array configuration is also marked on.  The error bar on the
number of sources $N$ in a band is $\sqrt{N}$.  The area within
$5^\circ$ of the Galactic plane is ignored.}
\label{figsurfnvss}
\end{figure}

\begin{figure}
\center
\epsfig{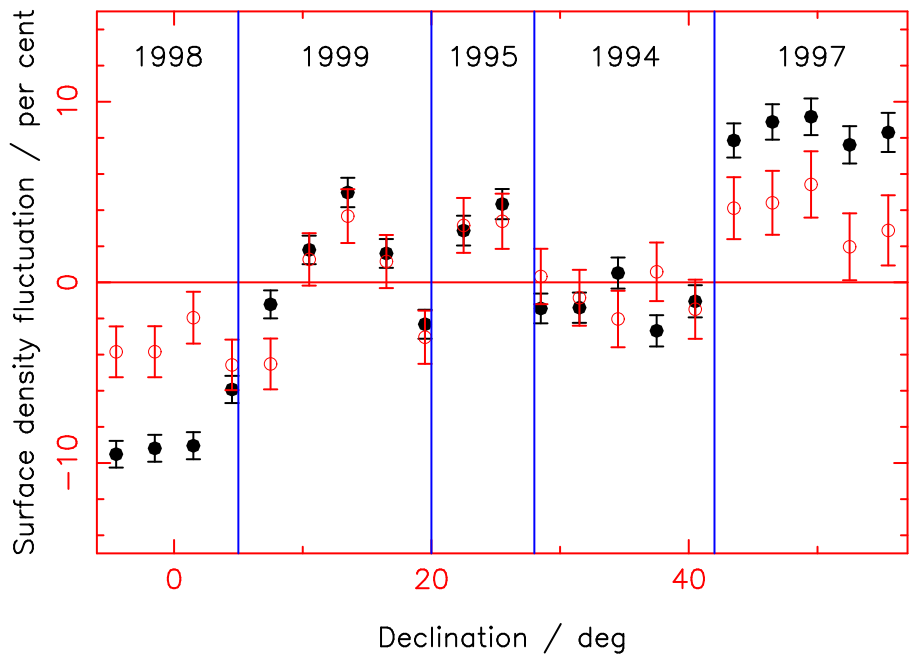}
\caption{Variations in FIRST source surface density as a function of
declination for sources with integrated flux densities above 2 mJy
(solid circles) and 10 mJy (open circles).  Sources have been binned
in declination bands of width $3^\circ$; the error bar on the number
of sources $N$ in a band is $\sqrt{N}$.  The approximate year in which
the observations in each declination region were made is also shown.}
\label{figsurffirst}
\end{figure}

Both the NVSS and FIRST surveys suffer from systematic fluctuations in
source surface density across the sky above flux-density thresholds at
which they are complete, affecting any attempt to quantify the
large-scale structure present.  Figures \ref{figsurfnvss} and
\ref{figsurffirst} illustrate these variations.  In both surveys the
magnitude of the effect depends on the flux-density threshold.  The
FIRST survey contains 10 per cent fluctuations in surface density over
the sky at the stated completeness limit $\fl = 2$ mJy, dropping to
below 5 per cent at 10 mJy.  The variations appear to correlate with
the different observing periods.  The NVSS suffers from 2 per cent
fluctuations at the completeness limit $\fl = 3.5$ mJy, with
significant density steps at the declinations where the array
configuration is changed from D to DnC, but these fluctuations have
become insignificant by 15 mJy.  These NVSS effects originate from
difficulties in compensating for the sparse $uv$-plane coverage of the
survey, constructed from ``snapshot'' radio observations, and are
purely declination-dependent: the projection of array baselines on the
sky changes with declination, whereas the data acquisition and
analysis software should be insensitive to the right ascension of
sources.  Note that the NVSS flux-density errors are dominated by a
constant additive bias that affects weak sources much more strongly
than powerful sources (unlike a multiplicative calibration error).

A varying source density $\sigma$ will spuriously enhance the measured
value of the angular correlation function $w(\theta)$.  This is
because the number of close pairs of galaxies in any region depends on
the local surface density ($DD \propto \overline{\sigma^2}$), but the
number of close pairs in the comparison random distribution depends on
the global average surface density ($RR \propto
(\overline{\sigma})^2$).  Systematic fluctuations mean that
$\overline{\sigma^2} > (\overline{\sigma})^2$, thus $w(\theta)$ is
increased.  The variance of counts-in-cells, as quantified by the
statistic $y(L)$ (\S \ref{secvar}), will also be increased: a spread
in the mean surface density across the cells will inevitably broaden
the overall probability distribution $P(N)$, which is constructed from
fluctuations about those means.  We can show that {\it on angular
scales less than those on which $\sigma$ is varying}, both $w(\theta)$
and $y(L)$ are subject to the same spurious constant offset
$\overline{\delta^2}$, where $\delta = (\sigma -
\overline{\sigma})/\overline{\sigma}$ is the surface overdensity.

To estimate the magnitude of this effect, take a simple toy model in
which a survey is divided into two equal areas between which there is
a fractional surface density shift $\epsilon$.  For this model,
$\overline{\delta^2} = \epsilon^2/4$.  Thus for example the expected
offsets on $w(\theta)$ are $\overline{\delta^2} \sim 2.5 \times
10^{-3}$ for FIRST sources above 2 mJy ($\epsilon \sim 0.1$) and
$\overline{\delta^2} \sim 2 \times 10^{-4}$ for NVSS sources above 3.5
mJy ($\epsilon \sim 0.03$).

In this study, analysis of the radio surveys was restricted to
flux-density ranges for which surface gradients were negligible.  The
alternative approach is to modulate the random comparison sets with
the same surface gradients as contained in the data.  The gradients
are not known in advance and must be measured from the data itself; it
was found that this could not be done sufficiently accurately to
subtract the offset completely.

\subsection{Multiple-component sources}
\label{secmult}

The large linear sizes and complex morphologies of radio sources mean
that a single radio galaxy can be resolved in a radio survey (and
appear in a survey catalogue) as two or more closely-separated
components of radio emission.  When investigating the clustering of
individual galaxies, these {\it multiple-component sources} will
produce spurious clustering at small separations.  There is no set of
criteria that will reliably distinguish multiple-component sources
from closely-separated independent galaxies; we instead choose to {\it
incorporate their effect into the fitted clustering model}.  As it
turns out, this effect may be successfully disentangled from that of
galaxy-galaxy clustering.

The influence of multiple-component sources on the angular correlation
function is quantified in detail in our analysis of the NVSS
$w(\theta)$ (Blake \& Wall 2002), in which it is stressed that {\it
even a tiny fraction of giant radio galaxies can substantially affect
clustering measurements at small angles.}  This is because a very
small number of close pairs determine the value of $w(\theta)$: we
find that we cannot neglect the effect of radio galaxies of size
$\theta$ until $\theta \sim 0.1^\circ$.  At angles $\theta <
0.1^\circ$, $w(\theta)$ effectively measures the {\it size
distribution} of radio galaxies; at $\theta > 0.1^\circ$, the {\it
clustering of individual galaxies} dominates the pair count.  This
reasoning is evidenced by the observed numbers and sizes of giant
radio sources (Lara et al. 2001) as well as a clear break in the
measured $w(\theta)$ at $\theta \approx 0.1^\circ$ (see Figure
\ref{figcorrnvss}).  Hence the influence of multiple-component sources
may be disentangled from that of galaxy clustering.

At angles where the clustering of individual galaxies dominates the
pair count, the fact that these galaxies may be split into multiple
radio sources is unimportant.  For if the mean number of radio
components per galaxy is $\overline{n}$, then the number of pair
separations $N_p$ at any angle is increased by a factor
$(\overline{n})^2$, and the source surface density $\sigma$ is
increased by a factor $\overline{n}$.  As $N_p \propto \sigma^2 \times
(1 + w)$, the measured correlation function is unaffected.

The effect of multiple-component sources on the moments of
counts-in-cells is modelled in Section \ref{seccellmult}.

\section{The radio galaxy angular correlation function}
\label{seccorr}

\subsection{Measurement of the NVSS angular correlation function}
\label{seccorrnvss}

\begin{figure}
\center
\epsfig{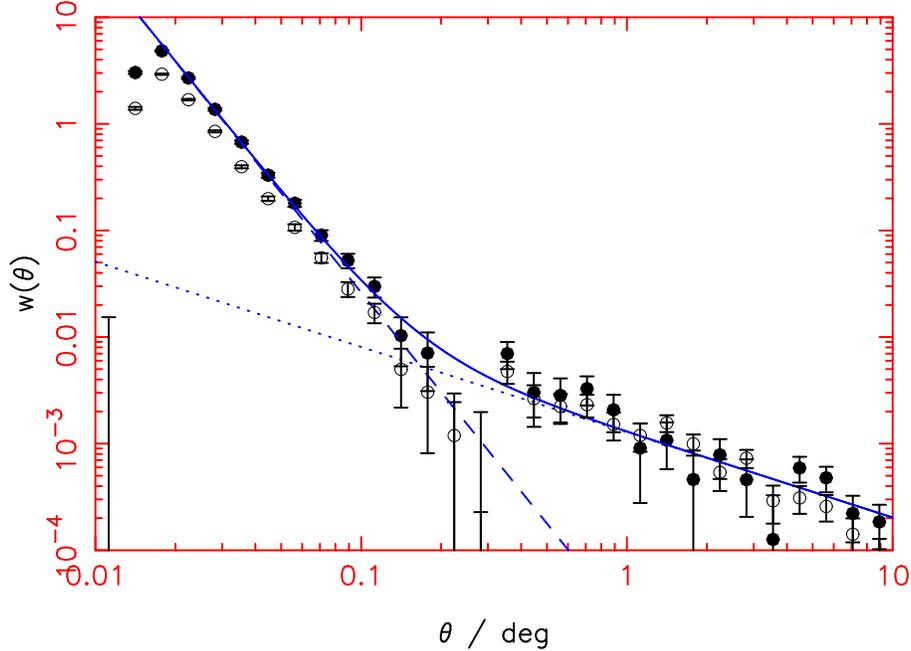}
\caption{Measurement of the NVSS angular correlation function for
flux-density thresholds $\fl = 20$ mJy (solid circles) and 10 mJy
(open circles).  The best-fitting sum of two power-laws for the 20 mJy
data is overplotted (as are the individual power-laws).  The amplitude
of the small-angle power-law, which is due to multiple-component
sources, decreases with flux-density threshold owing to the increasing
surface density (see Blake \& Wall 2002).  The parameters of the
large-angle power-law, which is due to galaxy clustering, are
independent of flux-density threshold.}
\label{figcorrnvss}
\end{figure}

Our measurements of the NVSS $w(\theta)$ for different flux-density
thresholds are described in Blake \& Wall (2002); the results for 10
mJy and 20 mJy are compared in Figure \ref{figcorrnvss}.  The angular
correlation function at all thresholds can be fit by a sum of two
power-laws.  A convincing interpretation is that the steep small-angle
power-law is created by multiple-component radio sources and hence
indicates their size distribution, whereas the shallow large-angle
power-law describes the clustering between different radio galaxies.

Table \ref{tabcorrnvss} displays the results of fitting the function
$w(\theta) = A \, \theta^{-\alpha} + B \, \theta^{-\beta}$ to the
measurements for flux-density thresholds $\fl = 10, 15, 20$ mJy.  The
fits were performed to angles $\theta > 0.02^\circ$, safely above the
resolution limit of the NVSS ($\theta_{{\rm res}} = 0.0125^\circ$).
Figure \ref{figcorrnvss} reveals a fall-off in $w(\theta)$ with
decreasing $\theta$ in the lowest separation bins.  This is not real,
but is a signal-to-noise problem caused by the failure of the survey
to resolve weak double sources with separations slightly greater than
the beam-width.  As $\theta$ increases, a rapidly increasing fraction
of doubles of size $\theta$ can be successfully resolved.

\begin{table}
\center
\caption{The best-fitting amplitudes and slopes of the double
power-law model $w(\theta) = A \, \theta^{-\alpha} + B \,
\theta^{-\beta}$ for the NVSS angular correlation function at
different flux-density thresholds; $\theta$ is measured in degrees.
The best fit is obtained by minimizing the $\chi^2$ statistic.  The
errors in the parameters are derived by varying each in turn from the
best-fitting combination (keeping the others fixed) and determining
the variation for which $\Delta \chi^2 = 1$, the appropriate 1-sigma
increment when varying one fitted parameter.  The value of $\chi^2$ is
not strictly meaningful given the possible correlations between
adjacent separation bins.  The reduced $\chi^2$ of the best fit and
the number of sources $n$ analyzed at each flux-density threshold are
also indicated.}
\label{tabcorrnvss}
\begin{tabular}{ccccccc}
\hline
$\fl$ & $n$ & $A$ / $\times 10^{-3}$ & $\alpha$ & $B$ /
$\times 10^{-5}$ & $\beta$ & $\chi_{{\rm red}}^2$ \\
\hline
$> 20$ & 292,962 & $1.22 \pm 0.15$ & $0.76 \pm 0.08$ & $2.49 \pm 0.09$
& $3.05 \pm 0.01$ & 1.15 \\
$> 15$ & 375,697 & $1.10 \pm 0.12$ & $0.84 \pm 0.07$ & $1.62 \pm 0.03$
& $3.11 \pm 0.01$ & 2.25 \\
$> 10$ & 522,341 & $1.08 \pm 0.09$ & $0.83 \pm 0.05$ & $0.95 \pm 0.02$
& $3.18 \pm 0.01$ & 2.24 \\
\hline
\end{tabular}
\end{table}

For all flux-density thresholds, the slope of the clustering power-law
is consistent with $\alpha = 0.8$ (in agreement with other classes of
objects) with an amplitude $A \approx 1 \times 10^{-3}$ (with $\theta$
in degrees).  Blake \& Wall (2002) describe a preliminary analysis of
the implications for spatial clustering.

\subsection{Comparison with the FIRST angular correlation function}
\label{seccorrfirst}

\begin{figure}
\center
\epsfig{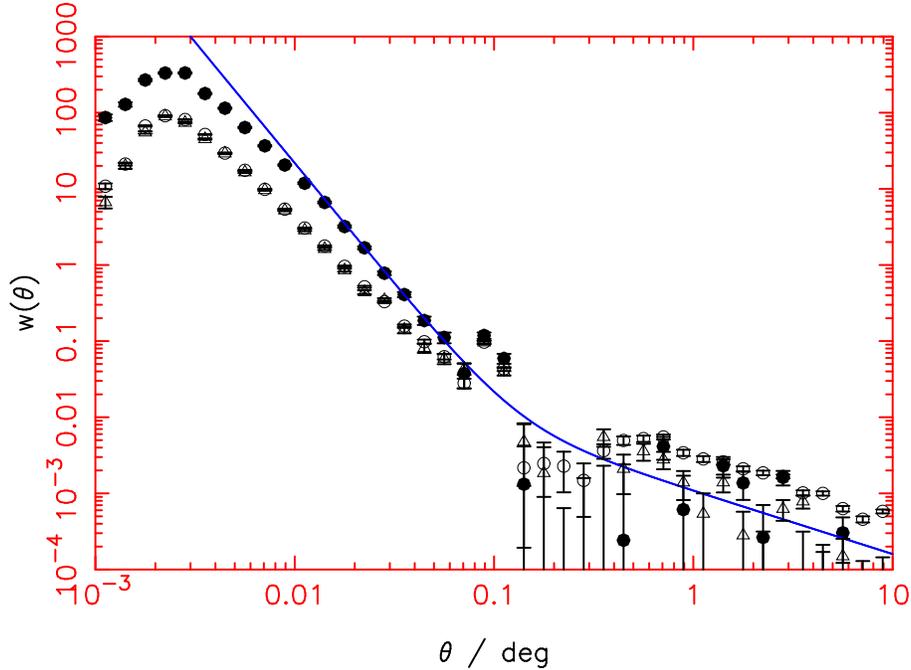}
\caption{Measurements of $w(\theta)$ from the FIRST catalogue.  This
Figure plots the whole-survey measurements at 10 mJy (solid circles)
and 2 mJy (open circles) and the $42^\circ < \delta < 57^\circ$
measurement at 2 mJy (triangles).  The solid line is the best-fitting
NVSS sum of two power-laws at 10 mJy (see Table \ref{tabcorrnvss}).
The smaller angular resolution of the FIRST survey permits us to
explore $w(\theta)$ down to $\theta \approx 0.001^\circ \approx 5$
arcsec.}
\label{figcorrfirst}
\end{figure}

For comparison, we measured $w(\theta)$ from the FIRST catalogue over
the region $123^\circ < \alpha < 247^\circ$, $5^\circ < \delta <
58^\circ$ for all objects above flux-density thresholds $\fl = 10$ mJy
and 2 mJy.  As a precautionary measure, we placed circular masks of
radius $0.5^\circ$ around all sources with $\fl > 1$ Jy; this left
respectively 88,873 and 305,872 objects at the two thresholds.  Figure
\ref{figcorrfirst} displays the results, compared to the best-fitting
NVSS sum of two power-laws at 10 mJy.  The FIRST measurement at
$\theta \approx 0.1^\circ$ is known to be contaminated with sidelobes
as described by Cress et al. (1996).

The FIRST 10 mJy and NVSS 10 mJy measurements agree well at small
angles, in the regime of multiple-component sources.  At bigger
angles, the large FIRST error bars mean that $w(\theta)$ is poorly
constrained, although the measurement is consistent with the NVSS
result ($\chi_{{\rm red}}^2 = 1.84$ for $\theta > 0.02^\circ$
excluding the two points at $\theta \approx 0.1^\circ$).  The FIRST 2
mJy $w(\theta)$ measurement has much smaller errors but is offset by
the source surface density gradients described in Section
\ref{secsurf}.  To verify this, we re-ran the 2 mJy analysis for the
more uniform declination region $42^\circ < \delta < 57^\circ$,
prompted by Figure \ref{figsurffirst}, using right ascension range
$107^\circ < \alpha < 263^\circ$.  The re-determined FIRST 2 mJy
$w(\theta)$ is in much better agreement with the NVSS result at large
angles (see Figure \ref{figcorrfirst}).  The amplitude of the
small-angle FIRST $w(\theta)$ drops between 10 mJy and 2 mJy due to
the increased surface density, consistent with the hypothesis that
$w(\theta)$ in this regime is entirely governed by multiple-component
sources (see Blake \& Wall 2002).

\section{Comparison with the radio galaxy counts-in-cells}
\label{seccell}

We also performed a counts-in-cells analysis of the radio surveys.
Our motivation was twofold: to quantify the clustering imprint with an
independent statistic to compare with $w(\theta)$, and to make contact
with previous work (Magliocchetti et al. 1998, 1999).

\subsection{Relation of counts-in-cells variance to $w(\theta)$}
\label{secvar}

Consider an unclustered distribution of sources, distributed randomly
and independently with surface density $\sigma$.  The expected number
of sources in a cell of area $S$ is $<N> = \sigma \times S$.  The
expected probability distribution $P(N)$ is the Poisson distribution
with mean $<N>$ and variance $<N>$.  We define the following statistic
to quantify the increased variance of a clustered distribution:
\[ y = \frac{\mu_2 - \overline{N}}{(\overline{N})^2} \]
Hence $<y> = 0$ for no clustering, as $<\mu_2> = <\overline{N}> =
<N>$.  (Actually, as there is statistical error in the denominator,
$y$ contains a slight bias which may be neglected).  We can show
(Peebles 1980, equation 36.6) that for a given $w(\theta)$, the
expected value of $y$ is
\begin{equation}
<y> = \frac{\int_{{\rm cell}} \int_{{\rm cell}} w(\theta) \, dS_1 \,
dS_2}{S^2} = \int w(\theta) \, \frac{dG_p}{d\theta} d\theta
\label{eqy1}
\end{equation}
where in the final expression, $dG_p$ is the fraction of all pairs of
area elements within the cell lying in the separation range $\theta
\rightarrow \theta + d\theta$ (using the notation of Landy \& Szalay
1993).  $dG_p$ may be calculated analytically for simple geometries:

\begin{itemize}
\item For a square of side $L$, putting $x = \theta/L$,
\begin{equation}
\frac{dG_p}{dx} = \cases {2x \, (\pi - 4x + x^2) & $0 \le x \le 1$ \cr
2x \, [\pi - 2 + 4\sqrt{x^2-1} - 4\arccos{(1/x)} - x^2] & $1 \le x \le
\sqrt{2}$ \cr }
\label{eqdgpsq}
\end{equation}
\item For a circular cell of diameter $L$, putting $x = \theta/L$,
\begin{equation}
\frac{dG_p}{dx} = \frac{16 \, x}{\pi} \left( \arccos{(x)} - x
\sqrt{1-x^2} \right)
\label{eqdgpcirc}
\end{equation}
\end{itemize}

We now calculate $<y>$ for a power-law angular correlation function.
If the survey has angular resolution $\theta_{{\rm res}}$ then
\[ w(\theta) = \cases {-1 & $\theta < \theta_{{\rm res}}$ \cr
(\theta/\theta_0)^{-\alpha} & $\theta > \theta_{{\rm res}}$ \cr } \]
Hence from equation \ref{eqy1},
\[ <y> = - \int_0^{\theta_{{\rm res}}} \frac{dG_p}{d\theta} \, d\theta
+ \int_{\theta_{{\rm res}}}^{\theta_{{\rm max}}} \left(
\frac{\theta}{\theta_0} \right)^{-\alpha} \, \frac{dG_p}{d\theta} \,
d\theta \]
where $\theta_{{\rm max}} = L$ for a circular cell of diameter $L$ and
$\theta_{{\rm max}} = L\sqrt{2}$ for a square cell of side $L$.  To
unveil the clustering pattern we measure the variation of $<y>$ with
cell size $L$, for fixed cell shape.  This is elegantly done by the
substitution $x = \theta/L$:
\begin{equation}
<y(L)> = - \int_0^{\theta_{{\rm res}}/L} \frac{dG_p}{dx} \, dx +
\left( \frac{\theta_0}{L} \right)^\alpha \int_{\theta_{{\rm
res}}/L}^{x_{{\rm max}}} x^{-\alpha} \, \frac{dG_p}{dx} \, dx
\label{eqy2}
\end{equation}
The quantity $dG_p/dx$ is determined purely by the cell shape, not
size (see equations \ref{eqdgpsq} and \ref{eqdgpcirc}).  For small
$x$, $dG_p/dx = kx$ independently of cell shape ($k = 8$ for circular
cells and $k = 2\pi$ for square cells).  Assuming that $\theta_{{\rm
res}} \ll L$, this solves the first integral of equation \ref{eqy2}:
\[ <y(L)> = - \frac{k}{2} \left( \frac{\theta_{{\rm res}}}{L}
\right)^2 + \left( \frac{\theta_0}{L} \right)^\alpha
\int_{\theta_{{\rm res}}/L}^{x_{{\rm max}}} x^{-\alpha} \,
\frac{dG_p}{dx} \, dx \]

\begin{itemize}
\item If the slope of $w(\theta)$ is steep enough ($\alpha > 2$), the
non-Poisson clustering will be dominated by close pairs ($x \ll 1$)
and for all cell shapes we can neglect edge effects by assuming that
$dG_p/dx = kx$ and $x_{{\rm max}} = \infty$.  This allows us to solve
the second integral of equation \ref{eqy2} and obtain:
\begin{equation}
<y(L)> = k \left( \frac{\theta_{{\rm res}}}{L} \right)^2 \left[
\frac{1}{(\alpha-2)} \left( \frac{\theta_0}{\theta_{{\rm res}}}
\right)^\alpha - \frac{1}{2} \right]
\label{eqysteep}
\end{equation}
\item Otherwise, cell shape is important and the exact solution is
\begin{eqnarray}
<y(L)> &=& k \left( \frac{\theta_{{\rm res}}}{L} \right)^2 \left[
\frac{1}{(\alpha-2)} \left( \frac{\theta_0}{\theta_{{\rm res}}}
\right)^\alpha - \frac{1}{2} \right] \nonumber \\ &+& \left(
\frac{\theta_0}{L} \right)^\alpha \left[ \int_{x_{{\rm min}}}^{\infty}
x^{-\alpha} \frac{dG_p}{dx} dx - \frac{k}{(\alpha-2) x_{{\rm
min}}^{\alpha-2}} \right]
\label{eqy3}
\end{eqnarray}
where $x_{{\rm min}} = \theta_{{\rm lim}}/L_{{\rm min}}$, and the
integral must be solved numerically.
\end{itemize}

Thus equation \ref{eqy3} shows that in general, $<y>$ has a variation
with cell size of the form
\begin{equation}
<y(L)> = a \, L^{-2} + b \, L^{-\alpha}
\label{eqy4}
\end{equation}
where $a$ and $b$ are constants.  The limited angular resolution
$\theta_{{\rm res}}$ reduces the variance: the existence of an object
in a cell limits the available space in which other objects can
appear.  This effect varies with cell size because it depends on the
scale of the resolution relative to the cell size.  If a survey has
sharp enough angular resolution then the first term of equation
\ref{eqy4} may be neglected (if $\alpha < 2$), and $<y> \propto
L^{-\alpha}$.  This is the case for the FIRST survey but not for the
NVSS; thus equation 9 of Magliocchetti et al. (1998) only contains the
second term of our equation \ref{eqy4}.

The following statistic is commonly defined to characterize the
departure of the skewness from a Poisson distribution:
\begin{equation}
z = \frac{\mu_3 - 3\mu_2 + 2\overline{N}}{(\overline{N})^3}
\label{eqskew}
\end{equation}
This has the expectation value
\[ <z> = \frac{\int_{{\rm cell}} \int_{{\rm cell}} \int_{{\rm cell}}
W(\theta_{12},\theta_{13},\theta_{23}) \, dS_1 \, dS_2 \, dS_3}{S^3}
\]
where $W(\theta_{12},\theta_{13},\theta_{23})$ is the three-point
angular correlation function.  The significance of the skewness is
summarized in Section \ref{secquan} above.

\subsection{Effect of multiple-component sources on the
counts-in-cells moments}
\label{seccellmult}

The existence of multiple-component sources increases the moments of
counts-in-cells.  This is because the fraction of radio sources within
a cell that are split into multiple components varies from cell to
cell, which acts to broaden the probability distribution of
counts-in-cells.  The simplest model is to suppose that a fraction $e$
of the galaxies are double sources (i.e. two sources appearing at the
same location in space) and a fraction $f$ are triple sources.  It can
be shown (Blake 2002) that the expected offsets in the variance and
skewness statistics are
\begin{equation}
\Delta y = \frac{1}{\sigma \, S} \left( \frac{2e + 6f}{1 + e + 2f}
\right)
\label{eqymult}
\end{equation}
\begin{equation}
\Delta z = \frac{1}{(\sigma \, S)^2} \left( \frac{6f}{1 + e + 2f}
\right)
\label{eqzmult}
\end{equation}
where $S$ is the cell area and $\sigma$ is the surface density of all
components (i.e. catalogue entries).  Thus skewness is sensitive only
to triple sources.

This simple model neglects the fact that the components of a radio
galaxy have a range of non-zero separations, but this should only
matter if the cell size is not much greater than the maximum component
separation ($\sim 0.1^\circ$).  A more sophisticated treatment is to
model the separation distribution by the effective (small-angle)
$w(\theta)$ of Figure \ref{figcorrnvss}.  We can compute the effect on
the variance statistic $y$ using equation \ref{eqy4}, thus the general
expression for $y(L)$ can be modified to
\[ <y(L)> = a \, L^{-2} + b \, L^{-\alpha} + c \, L^{-\beta} \]
where $c$ is a constant and $\alpha$ and $\beta$ are respectively the
slopes of the shallow (galaxy clustering) and steep (multiple
component) $w(\theta)$ power-laws.

It is easy to show that this more sophisticated model reduces in the
appropriate limit to the more simple treatment initially outlined.  As
component separations tend to zero, the slope $\beta$ of the effective
$w(\theta)$ becomes large and we can use the ``steep clustering''
approximation of equation \ref{eqysteep}, which reproduces the
dependence $y \propto L^{-2}$ of equation \ref{eqymult}.

\subsection{Error on the counts-in-cells moments}

Variance and skewness measurements are subject to statistical error
due to averaging over a finite number of cells $N_c$.  Calculating the
standard error on the statistics $y$ and $z$ in the case of a random
(unclustered) distribution yields
\begin{equation}
\sigma_y = \sqrt{\frac{2}{N_c \, (\overline{N})^2}}
\label{eqyerr}
\end{equation}
\begin{equation}
\sigma_z = \sqrt{\frac{6}{N_c \, (\overline{N})^3}}
\label{eqzerr}
\end{equation}
The probability distribution of the clustered data does not depart
greatly from a Poisson distribution ($y \ll 1$, $z \ll 1$), so that
these expressions are very good approximations to the actual
statistical errors.

\subsection{Measuring the counts-in-cells moments from a real survey}

To measure the counts-in-cells we used the simple technique of
defining a grid of touching circular cells of diameter $L$ on the sky
and counting the number of sources that fall inside each cell.  This
only utilises a fraction $\pi/4$ of the available area, but the cell
shape is constant over the sky.  We note that the methodology of
counts-in-cells was revolutionized by Szapudi (1998) who showed that
it was valid to throw a very large number of randomly-placed cells
over the sky, heavily oversampling the survey area.  We prefer the
former, simpler approach for a first investigation aiming to show
consistency with the $w(\theta)$ analysis.

Surveys do not encompass the whole sky: there are boundaries and
masked regions.  Hence some cells in the grid are partially filled,
the $i$th cell having fraction of useful area $f_i$ (say).  To
determine $f_i$ for each cell we populated the sky with random points
subject to the same boundaries and masks as the real survey.  Counting
the number of random points that fall in each cell accurately measures
the useful area.  We then boosted the data count in the $i$th cell by
a factor $1/f_i$, unless $f_i$ was less than a threshold $f_{{\rm
rej}} = 0.75$ in which case we rejected the cell.  To measure the
factors $f_i$ accurately enough it is essential to average over a
sufficiently large number of random realizations that statistical
noise does not dominate.  Let there be $m$ random sets, each with the
same surface density as the survey.  The lower limit on $m$ is
determined by the following considerations:
\begin{itemize}
\item{The correction of cell counts by factors $1/f_i$ creates extra
variance in the counts-in-cells (because all cells are corrected,
whether they are partially filled or not).  This extra systematic
variance must be much smaller than the statistical error of equation
\ref{eqyerr}.  This condition is equivalent to
\[ m \gg m_1 = \sqrt{\frac{N_c}{2}} \left( 1 + \frac{1}{\overline{N}}
\right) \]
}
\item{The cell areas must be determined precisely enough that a
negligible fraction of ``unspoilt'' cells are rejected with $f_i <
f_{{\rm rej}}$.  This condition is equivalent to
\[ m \gg m_2 = \frac{1}{\overline{N} \, (1 - f_{{\rm rej}})^2} \]
}
\end{itemize}

When evaluating the moments of the counts-in-cells distribution we
assume that all cells are populated independently.  This is not
strictly true given that clustered sources have correlated positions,
but the assumption should be a very good approximation if the cells
are large enough.  With this consideration in mind, we adopted a
minimum cell size $L_{{\rm min}} = 0.3^\circ$ for our analysis; below
this cell size the average number of sources contained in a cell drops
below $\overline{N} = 1$ at the relevant flux-density thresholds.

\subsection{Measurement of the NVSS counts-in-cells moments}
\label{seccellnvss}

Using these methods, we measured the counts-in-cells variance
statistic $y(L)$ from the NVSS for circular cells of diameters
$0.3^\circ < L < 10^\circ$ for flux-density thresholds $\fl = 10$ mJy
and 20 mJy.  As described in Blake \& Wall (2002), we masked out all
NVSS catalogue entries within $5^\circ$ of the Galactic plane and also
within 22 additional masked regions around radio sources that appear
in the NVSS catalogue as a large number of separate elliptical
Gaussians.  As discussed in Section \ref{seccorrnvss}, the difficulty
in resolving faint doubles with separations just greater than the
beam-width causes some variation in the effective survey angular
resolution, as evidenced by the fall-off in $w(\theta)$ with
decreasing $\theta$ in the lowest separation bins of Figures
\ref{figcorrnvss} and \ref{figcorrfirst}.  To ensure a consistent
value of $\theta_{{\rm res}}$ for the prediction of the
counts-in-cells variance from $w(\theta)$, we ran a linking algorithm
that combined together all pairs of sources with angular separations
less than 1 arcmin, thus artificially establishing $\theta_{{\rm res}}
= 1$ arcmin.  This left 289,981 NVSS catalogue entries above 20 mJy
and 516,782 entries above 10 mJy.

\begin{figure}
\center
\epsfig{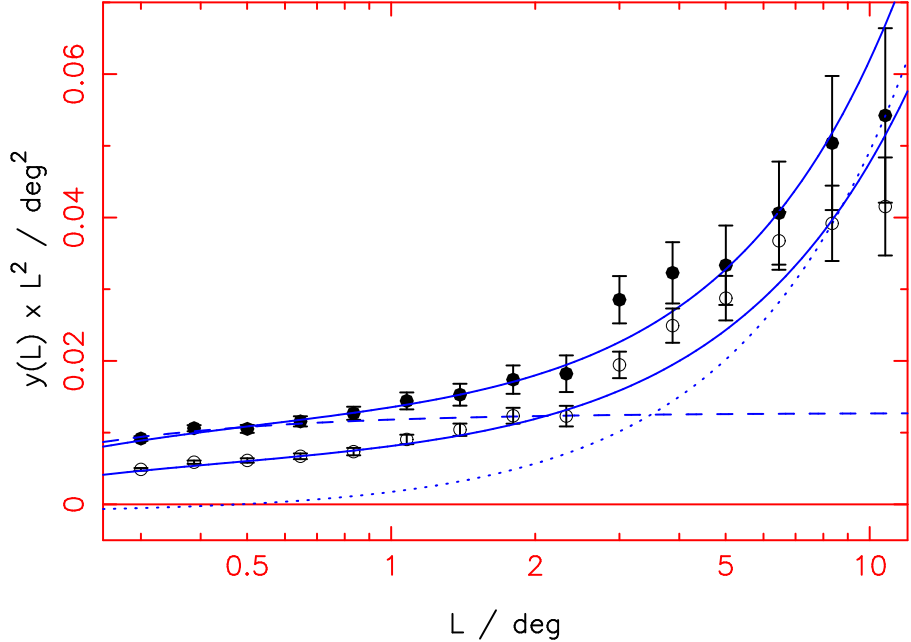}
\caption{The NVSS counts-in-cells variance statistic $y(L)$ is plotted
for thresholds 20 mJy (solid circles) and 10 mJy (open circles)
together with the predictions of the double power-law $w(\theta)$
model at 20 mJy and 10 mJy (the solid lines).  The dashed and dotted
lines show the separate contributions to $y(L)$ at 20 mJy of the steep
(multiple-component) $w(\theta)$ and shallow (galaxy clustering)
$w(\theta)$.  The angular correlation function predictions provide a
good fit to the measurements of the variance, demonstrating the
consistency of these independent methods of quantifying angular
structure in the NVSS.}
\label{figvarnvss}
\end{figure}

The variance measurements at the two flux-density thresholds are
plotted in Figure \ref{figvarnvss}, with error bars from equation
\ref{eqyerr}.  We also plot the predictions generated from the double
power-law model for $w(\theta)$ with best-fit coefficients at the
appropriate flux-density thresholds as listed in Table
\ref{tabcorrnvss}.  The counts-in-cells variance is visually
consistent with the prediction of the measured angular correlation
function, verifying the agreement of these two independent methods of
quantifying angular structure in the NVSS.  The $\chi^2$ statistics
between the data and the prediction are $\chi_{{\rm red}}^2 = 0.69$ at
20 mJy and $\chi_{{\rm red}}^2 = 1.64$ at 10 mJy.  However, the value
of $\chi^2$ is not strictly meaningful because the variances for
different cell sizes are not independent.  Note that
multiple-component sources produce an approximately constant offset in
$y \, L^2$ in Figure \ref{figvarnvss}.  The variance statistic for the
10 mJy threshold falls below that for the 20 mJy measurement because
this offset is proportional to $1/\sigma$, where $\sigma$ is the
source surface density (see equation \ref{eqymult}).

The angular correlation function model used to generate the
predictions is a sum of two power-laws, representing respectively
multiple-component sources ($w(\theta) \propto \theta^{-3.1}$) and
galaxy clustering ($w(\theta) \propto \theta^{-0.8}$).  To assess the
contribution of each effect to the overall counts-in-cells variance we
separately converted each of the two 20 mJy power-laws to a variance
and plotted the result on Figure \ref{figvarnvss}.  For small cell
sizes ($L < 1^\circ$) the extra variance produced by
multiple-component sources dominates.  For larger cell sizes the extra
variance produced by galaxy clustering becomes increasingly important.
This type of transition is expected, as the number of extra pairs at
angle $\theta$ scales as $w(\theta) \times 2 \pi \theta \, d\theta$,
which varies as $\theta^{-2.1}$ for multiple-component sources and
$\theta^{+0.2}$ for pairs of galaxies.

\begin{figure}
\center
\epsfig{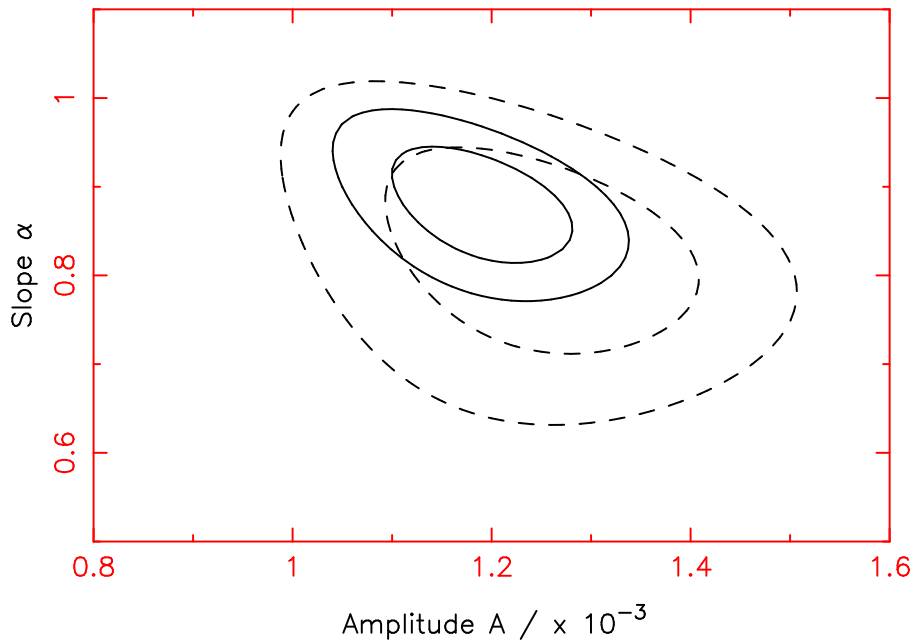}
\caption{Constraints on the clustering parameters $w(\theta) = A \,
\theta^{-\alpha}$ from counts-in-cells.  The angular correlation
function may be converted into a variance of counts-in-cells (Section
\ref{secvar}) and thereby compared with the counts-in-cells
measurements.  Multiple components are modelled as a second power-law
$w(\theta)$, the coefficients of which are held constant at their
best-fitting values (Table \ref{tabcorrnvss}).  The Figure shows
$1\sigma$ and $2\sigma$ contours in the space of $(A,\alpha)$ for
flux-density thresholds of 10 mJy (solid) and 20 mJy (dashed).  As the
plot is in the space of two varying parameters, these contours are
defined by $\chi^2$ increasing by respectively 2.30 and 6.17 from its
minimum, although as variance measurements for different cell sizes
are not independent, the value of $\chi^2$ is not strictly
meaningful.}
\label{figcellparam}
\end{figure}

By converting the measured $w(\theta)$ into an equivalent variance we
have shown that the angular correlation function and counts-in-cells
analyses are entirely consistent.  Alternatively, we can make an
independent determination of the angular correlation function
parameters by finding the best fit to $y(L)$.  We varied the amplitude
and slope of the clustering power-law $w(\theta) = A \,
\theta^{-\alpha}$, whilst keeping the multiple-component $w(\theta)$
parameters fixed at their best-fitting values from Table
\ref{tabcorrnvss}.  Thus for each point in the $(A,\alpha)$ grid we
obtained a model $y(L)$, which we compared with the measured $y(L)$
using the $\chi^2$ statistic.  The best-fitting clustering parameters
were in good agreement with those derived from the original
$w(\theta)$ analysis (see Figure \ref{figcellparam}).

\begin{figure}
\center
\epsfig{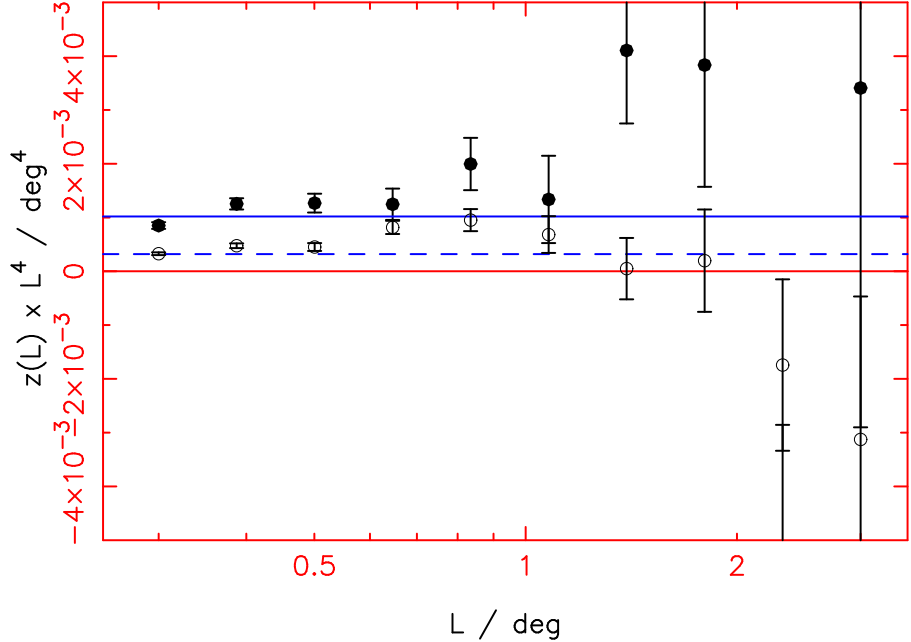}
\caption{Measurement of the NVSS skewness statistic $z(L)$ for
thresholds 20 mJy (solid circles) and 10 mJy (open circles).  The
prediction of the multiple-component model of equation \ref{eqzmult}
is also plotted for 20 mJy (solid line, $\sigma = 9.37$ deg$^{-2}$)
and 10 mJy (dashed line, $\sigma = 16.69$ deg$^{-2}$) assuming a
fraction of doubles $e = 0.07$ and triples $f = 0.01$.  The skewness
statistic becomes dominated by noise for cell sizes $L > 1^\circ$ due
to the decreasing number of cells contained in the grid.  It is most
convenient to plot $z \times L^4$ against $L$ to illustrate the
influence of multiple components.}
\label{figskewnvss}
\end{figure}

Figure \ref{figskewnvss} displays measurements of the skewness
statistic $z(L)$ (equation \ref{eqskew}) from the NVSS at 20 mJy and
10 mJy.  The result is a significant detection of skewness.  However,
the fact that $z$ scales roughly as $L^{-4}$ suggests that the
skewness is dominated by multiple-component sources, especially as the
amplitude of the skewness scales with surface density in accordance
with equation \ref{eqzmult}.  On Figure \ref{figskewnvss} we plot the
multiple-component skewness predictions at 10 mJy and 20 mJy assuming
that 7 per cent of radio galaxies are doubles (see Blake \& Wall 2002)
and 1 per cent are triples (i.e. $e = 0.07$, $f = 0.01$).  This
fraction of triple sources is reasonable; analyzing NVSS sources into
groups using link-length $\theta_{{\rm link}} = 5$ arcmin (a
reasonable approximation to the range of dominance of
multiple-component sources, see Figure \ref{figcorrnvss}) produces
groups of which 9.8 per cent are doubles and 1.5 per cent are triples.
These skewness predictions due to multiple components produce a fairly
good fit to the data.  The NVSS hence provides no convincing evidence
for cosmological skewness.

\subsection{Comparison with the FIRST counts-in-cells variance}

We also derived counts-in-cells for the FIRST radio survey, previously
studied by Magliocchetti et al. (1998).  Our results are not directly
comparable to those of Magliocchetti et al. (1998) as we make no
attempt to combine multiple-component sources.  We analyzed the three
FIRST sub-samples used in the angular correlation function study of
Section \ref{seccorrfirst}.  To ensure a consistent survey angular
resolution we first ran the source-combining algorithm with
link-length $\theta_{{\rm res}} = 0.003^\circ$.

\begin{figure}
\center
\epsfig{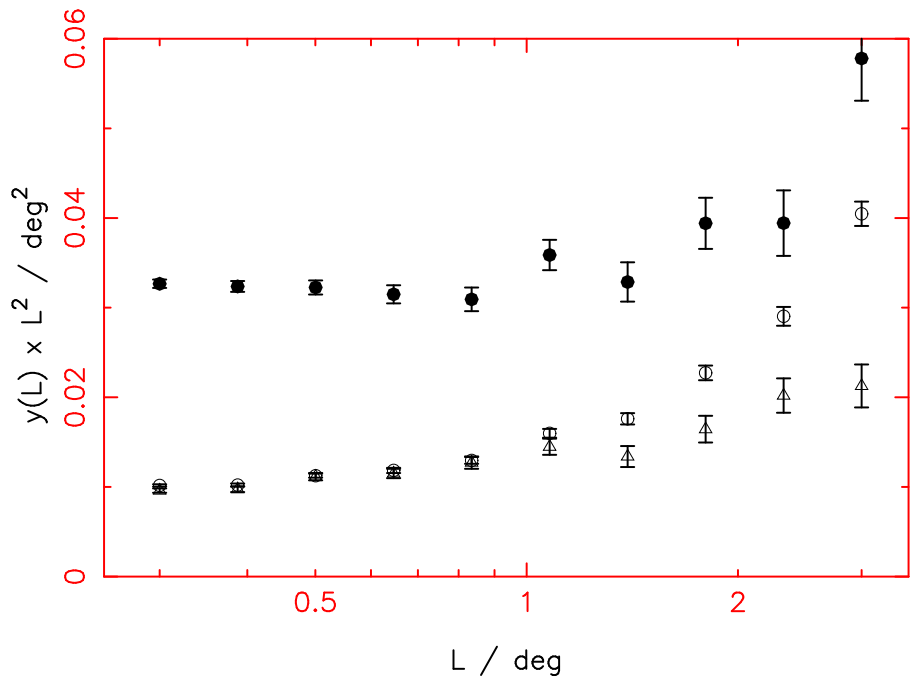}
\caption{The FIRST counts-in-cells variance $y(L)$ is plotted for the
three samples defined in the text: 10 mJy all-sky (solid circles), 2
mJy all-sky (open circles) and 2 mJy reduced area (triangles).  The
errors in the data points are determined using equation \ref{eqyerr}.}
\label{figcellfirst}
\end{figure}

The variance results for the three FIRST sub-samples are plotted in
Figure \ref{figcellfirst}.  In all samples there is a strong
contribution from multiple components ($y \, L^2 =$ constant) which
dominates at small cell sizes.  The amplitude of this contribution
varies as $1/\sigma$ where $\sigma$ is the source surface density
(equation \ref{eqymult}), which accounts for the overall shift between
the 10 mJy and 2 mJy samples (between which $\sigma$ varies by a
factor $\approx 3.4$).  Angular resolution effects are unimportant for
FIRST; hence there is no dip in the variance at small $L$.  Galaxy
clustering becomes important at higher $L$ ($y \propto L^{-\alpha}$).
The difference between the two samples at 2 mJy arises from the
surface density gradients present in the whole-sky sample.  Gradients
offset the counts-in-cells variance by $\Delta y =$ constant (Section
\ref{secsurf}), as is apparent from the increasing difference between
the plotted triangles and open circles in Figure \ref{figcellfirst}
(which plots $y \, L^2$ against $L$).  Comparing Figures
\ref{figcellfirst} and \ref{figvarnvss}, the greater angular
resolution of FIRST with respect to NVSS leads to an increased
abundance of multiple-component sources and thus a greater variance
signal for a given flux-density threshold.

\subsection{Comparison with previous work}

Our results differ from those of Magliocchetti et al. (1998) in two
respects:

\begin{enumerate}

\item Magliocchetti et al. (1998) reported a much steeper slope for
the correlation function, $\gamma = 2.5 \pm 0.1$ (where $w(\theta)
\propto \theta^{1-\gamma}$) compared to our NVSS measurement $\gamma =
1.83 \pm 0.05$ (Table \ref{tabcorrnvss}).  Magliocchetti et
al. adopted a combining algorithm for multiple components and assumed
that, after the operation of this algorithm, any remaining pairs were
independent radio galaxies.  However, no set of criteria can
unequivocally distinguish multiple components from independent
galaxies, and a small number of residual multiple-component sources
can have a dramatic effect on the clustering statistics on angular
scales up to several arc-minutes (see Blake \& Wall 2002).  As
demonstrated by Figure \ref{figcorrnvss}, multiple components are the
dominant provider of close pairs in the NVSS up to $\theta \approx
0.1^\circ$, whereas the fraction of closely-separated pairs combined
by Magliocchetti et al. becomes negligible by $0.02^\circ$.  Thus the
steep slope found by Magliocchetti et al. may be a manifestation of
the remaining multiple-component sources, and not of galaxy
clustering.  Furthermore, it is difficult to understand a steep slope
$\gamma = 2.5$ persisting to small angles, given that the $\gamma
\approx 1.8$ clustering law is obeyed by all other studied classes of
galaxy including local optically-selected galaxies (both spirals and
ellipticals, e.g. Loveday et al. 1995) and high-redshift QSOs (Croom
et al. 2001).

\item Magliocchetti et al. (1998) reported a cosmological skewness.
We suggest that this may also be produced by uncombined doubles.  The
relation $z \propto y^2$ used by Magliocchetti et al. as evidence for
the non-linear gravitational growth of perturbations (their equation
20) is also naturally produced in a model where multiple components
dominate -- it results from combining our equations \ref{eqymult} and
\ref{eqzmult}.  An alternative explanation lies in the different
flux-density limits (10 mJy for NVSS versus 3 mJy for FIRST).  The
FIRST sample may contain a non-negligible fraction of low-redshift
starburst galaxies which trace non-linear clustering and hence are a
source of skewness.

\end{enumerate}

\section{Conclusions}
\label{secconc}

We have quantified the angular clustering in the NVSS and FIRST radio
surveys using two independent methods: the two-point angular
correlation function and the variance on counts-in-cells.  Our results
may be summarized as follows:

\begin{enumerate}
\item The results of angular correlation function and counts-in-cells
analyses of the surveys are entirely consistent.
\item The larger area and greater number of sources in the NVSS yield
a much clearer description of the clustering imprint.  The correlation
function has two contributions: that due to multiple components of the
same galaxy, dominant at $\theta < 0.1^\circ$, and that due to
clustering between galaxies, which dominates at larger angles.  A
clear break in $w(\theta)$ is evident between these scales.  Both of
these contributions are needed to explain the observed variance on
counts-in-cells.
\item The clustering part of the correlation function has a slope
consistent with that measured in the optical regime, $w(\theta)
\propto \theta^{-0.8}$; this is confirmed by our counts-in-cells
measurements.
\item Both the NVSS and FIRST surveys suffer from systematic
fluctuations in source surface density at flux-density thresholds at
which they purport to be complete.
\end{enumerate}

Our work disagrees with some previous conclusions drawn from the FIRST
survey:

\begin{enumerate}
\item We find a galaxy correlation slope consistent with that measured
in the optical, $\gamma \approx 1.8$, in contrast to somewhat steeper
slopes reported in previous analyses.  These steeper slopes may have
been produced by residual multiple component radio sources.
\item The skewness reported by Magliocchetti et al. (1998) may also be
due to these residual double sources.
\end{enumerate}

This investigation has improved our understanding of the methodology
of angular clustering analyses for large-scale radio surveys, of
relevant observational effects present in such surveys, and of the
derived structural parameters.  There is now the basis to use these
surveys to derive three-dimensional information on the very largest
structural scales, adopting more powerful statistical methods in
conjunction with the redshift databases to be provided by surveys such
as 2dF and SDSS.

\section*{acknowledgments}

We thank Lance Miller and Steve Rawlings for helpful comments on
earlier drafts of this paper.  We especially acknowledge Manuela
Magliocchetti for extensive and helpful discussions.

\end{document}